\documentclass{article}[12pt]
\usepackage{amssymb, amsmath}
\usepackage[english]{babel}
\usepackage{amsthm}
\usepackage{amsfonts}
\usepackage{commath}
\usepackage{graphicx}
\usepackage{pdfpages}
\usepackage{multirow}
\usepackage{algorithm}
\usepackage{algorithmic}
\usepackage{amsfonts}
\usepackage{mathtools}
\usepackage{verbatim}
\usepackage{geometry}
\usepackage{multirow}
\usepackage{natbib}

\author{
  Changye Wu\thanks{CEREMADE, Universit\'e Paris-Dauphine PSL, {\sf wu@ceremade.dauphine.fr}}
  \and
  Christian Robert\thanks{Universit\'e Paris Dauphine PSL, and University of Warwick, {\sf xian@ceremade.dauphine.fr}}
}
\title{Parallelising MCMC via Random Forests}
\begin{document}
\maketitle
\begin{abstract}
\noindent For Bayesian computation in big data contexts, the divide-and-conquer
MCMC concept splits the whole data set into batches, runs MCMC algorithms
separately over each batch to produce samples of parameters, and combines them
to produce an approximation of the target distribution. In this article, we
embed random forests into this framework and use each
subposterior/partial-posterior as a proposal distribution to implement importance
sampling. Unlike the existing divide-and-conquer MCMC, our methods are based on
scaled subposteriors, whose scale factors are not necessarily restricted to
$1$ or to the number of subsets. Through several experiments, we show that our methods
work well with models ranging from Gaussian cases to strongly non-Gaussian
cases, and include model misspecification.
\end{abstract}


\section{Introduction}
Markov chain Monte Carlo (MCMC) algorithm, a generic sampling method, is
ubiquitous in modern  statistics, especially in Bayesian fields. MCMC
algorithms require only the evaluation of the target pointwise, up to a
multiple constant, in order to sample from it. In Bayesian analysis, the object
of main interest is the posterior, which is not in closed form in general,
and MCMC has become a standard tool in this domain. However, MCMC is
difficult to scale and its applications are limited when the observation size
is very large, for it needs to sweep over the entire
observations set in order to evaluate the likelihood function at each
iteration. Recently, many methods have been proposed to better scale MCMC algorithms for
big data sets and these can be roughly classified into two groups
\cite{bardenet2015markov}: divide-and-conquer methods and subsampling-based methods.\\
\\
For divide-and-conquer methods, one splits the whole data set into subsets,
runs MCMC over each subset to generate samples of parameters and combine these to produce
an approximation of the true posterior. Depending on how MCMC is handled over
the subsets, these methods can be further classified into two sub-categories. Let
$\mathcal{X}$ be the whole data set and $\mathcal{X}_1, \cdots, \mathcal{X}_K$
be the subsets. Denote $\pi_0$ by the prior distribution over the parameter
$\theta$. One approach (\citealp{neiswanger2013asymptotically}, \citealp{nemeth2017merging},
\citealp{scott2016bayes}, \citealp{wang2013parallelizing},
\citealp{wang2015parallelizing}) consists in running MCMC over 
\begin{equation*}
\pi_k(\theta|\mathcal{X}_k) \propto \left(\pi_0(\theta)\right)^{1/K}\prod_{x\in\mathcal{X}_k}p(x|\theta).
\end{equation*}
The other approach (\citealp{minsker2014scalable}, \citealp{srivastava2015wasp}) targets 
\begin{equation*}
\pi_k(\theta|\mathcal{X}_k) \propto \pi_0(\theta)\prod_{x\in\mathcal{X}_k}p(x|\theta)^{K}.
\end{equation*}
\\
For subsampling-based methods, one uses a partition of the whole data set to
estimate the MH acceptance ratio at each iteration in order to accelerate the
MCMC algorithms. These approaches can also be classified into two finer
classes: exact subsampling methods and approximate subsampling methods,
according to their outputs. Exact subsampling approaches typically require to
explore an augmented space and to treat the target distribution as its invariant
marginal distribution. One direction \citep{quiroz2016exact} is to take
advantage of pseudo-marginal MCMC \citep{andrieu2009pseudo} via constructing
unbiased estimators of point-wise evaluations of the target density with
subsets of the data. Another approach is to leverage the piecewise deterministic
Markov processes (\citealp{bierkens2017piecewise}, \citealp{bierkens2016zig},
\citealp{bouchard2017bouncy}, \citealp{davis1984piecewise}, \citealp{davis1993markov},
\citealp{fearnhead2016piecewise}, \citealp{sherlock2017discrete},
\citealp{vanetti2017piecewise}),  which take the targets as the marginal
distributions of their invariant distributions. Approximate subsampling
approaches aim at constructing an approximation of the target distributions.
One approach \citep{bardenet2014towards,bardenet2015markov} is to
determine the acceptance of the proposals with high probability using subsets
of the data. Another approach  \citep{chen2014stochastic,
ding2014bayesian,welling2011bayesian}, is based on direct
modifications of exact methods. The seminal work in this direction is
stochastic gradient Langevin dynamics (SGLD) \citep{welling2011bayesian}.\\
\\
In this article, we propose two methods to scale MCMC algorithms, which are based on divide-and-conquer principles. However, unlike the former divide-and-conquer approaches, we run MCMC over 
 \begin{equation*}
\pi_k(\theta|\mathcal{X}_k) \propto \left\{\left(\pi_0(\theta)\right)^{1/K}\prod_{x\in\mathcal{X}_k}p(x|\theta)\right\}^{\lambda},
\end{equation*}
where $\lambda$ is not necessarily restricted to $1$ or $K$. Further, we use
random forests \citep{breiman2001random} to learn approximations of the
subposteriors and we take advantage of them to approximate the true posterior by
an additional MCMC or importance sampling step. Section 2 describes the proposed
methods in details and several numerical examples are presented in section 3.
In section 4, we discuss the limitations of these methods and conclude the paper. 

\section{Methodology}
\noindent Denote by $\mathcal{X} = \left\{x_1, x_2, \cdots, x_N\right\}$ the
whole data set of observations, where $x_i\sim p_{\theta}(\cdot)$ i.i.d. and
$\theta\in\Theta\subset \mathbb{R}^d$. Let $\pi_0(\theta)$ denote the prior
distribution on the parameter space $\Theta$. Splitting the whole data set
$\mathcal{X}$ into subsets $\mathcal{X}_1,\cdots, \mathcal{X}_K$, each with
same size $m = \frac{N}{K}$, the target of interest is the posterior distribution:
\begin{equation*}
\pi(\theta|\mathcal{X}) \propto \pi_0(\theta)\prod_{i=1}^Np(x_i|\theta)
\end{equation*}
For each subset $\mathcal{X}_k$, $k=1, \cdots, K$, we define the $\lambda_k$-subposterior as:
\begin{equation*}
\pi_k^{\lambda_k}(\theta|\mathcal{X}_k)=\frac{\left(\gamma_k(\theta|\mathcal{X}_k)\right)^{\lambda_k}}{Z_{k,\lambda_k}}, \quad \gamma_k(\theta|\mathcal{X}_k) =  \pi_0(\theta)^{\frac{1}{K}}\prod_{x\in\mathcal{X}_k}p(x|\theta),
\end{equation*}
where $Z_{k,\lambda_k}$ is the normalising constant of $\left(\gamma_k(\theta|\mathcal{X}_k)\right)^{\lambda_k}$. \\
\\
In the divide-and-conquer paradigm, one applies MCMC algorithms on each
subposterior, generates samples from them and combines these samples to
approximate the true posterior distribution. Existing divide-and-conquer MCMC
methods treat these $\lambda_k$'s in two possible ways: one is to set $\lambda_k=1$ for
all $k=1,\cdots, K$, in e.g. consensus Monte Carlo and Weierstrass  sampler; the
other is to set $\lambda_k = K$, in e.g. WASP \citep{srivastava2015wasp},
M-posterior \citep{minsker2014scalable}. By contrasat, in our method, the choice of
$\lambda_k$ is not restricted to these two options.  Besides, the above mentioned
methods, except for the one proposed by \cite{nemeth2017merging}, ignore a valuable byproduct ---
namely the value of the target distribution at the proposed sample points --- of an MCMC
algorithm. However, when compared with \cite{nemeth2017merging}, our methods 
differ in two aspects: one is to use the scaled subposterior, the other is that
the learning algorithm is cheaper at both training and prediction stages.\\
\\
In our method, we embed a regression procedure into a divide-and-conquer perspective
in order to propose approximations of the subposterior density functions and use them to
produce an approximation of the true posterior. Specifically, we run MCMC
algorithms with MH steps, such as MCMC, HMC, MALA, over
$\pi_k^{\lambda_k}(\theta|\mathcal{X}_k)$ to obtain samples $\{\theta_1^k,
\theta_2^k, \cdots, \theta_T^k\}$. Considering the construction of MH
acceptance ratios, we just need to evaluate
$\log{\gamma_k(\theta|\mathcal{X}_k)}$ pointwise, instead of computing the harder target,
$\pi_k^{\lambda_k}(\theta|\mathcal{X}_k)$, in order to bypass the derivation of the normalising
constant, a common numerical problem in computation. As a byproduct, we can get the
evaluations of $\log{\gamma_k(\theta|\mathcal{X}_k)}$ at certain points
$\{\vartheta_1^{k}, \vartheta_1^{k}, \cdots, \vartheta_{T_k}^k\}$, which are
the proposed values in MCMC algorithms. Running random forests or other regression
machine learning algorithms on
$$\left\{\bigg(\vartheta_1^k,
\log\gamma_k(\vartheta_1^k|\mathcal{X}_k)\bigg), \bigg(\vartheta_2^k,
\log\gamma_k(\vartheta_2^k|\mathcal{X}_k)\bigg), \cdots,
\left(\vartheta_{T_k}^k,
\log\gamma_k(\vartheta_{T_k}^k|\mathcal{X}_k)\right)\right\}$$
provides an estimator, $f_k$, of $\log\{\gamma_k(\theta|\mathcal{X}_k)\}$. The uses of such
$f_k$'s, $k=1, \cdots, K$, are double: one is to approximate
$\pi(\theta|\mathcal{X})$ with 
$$f(\theta) = \displaystyle{\exp\left\{\sum_{k=1}^Kf_k(\theta)\right\}}$$
and to run an additional MCMC over $f$ to obtain samples which are regarded as an
approximation of the posterior distribution;
the other usage is to approximate $\pi(\theta|\mathcal{X})$ with $f(\theta)$
and apply importance sampling algorithm on $f$ with proposal distribution
$\gamma_k^{\lambda_k}(\theta|\mathcal{X}_k)$.\\
\\
$\textbf{Remark 1:}$ \textbf{(Scale factors)} The scale factor $\lambda_k$ is
used to control the uncertainty in the subposterior. Roughly speaking, it controls
the range of the region from which the embedded regression algorithm learns
each subposterior. When the scale factor is too large, for instance, when $\lambda_k
= 100K, k=1,\cdots, K$, each scaled subposterior has small uncertainty, which
may lead to the resulting MCMC samples not overlapping with one another. In the 
event this happens, the approximation $f_k$ cannot provide useful information on the region where
$\gamma_j(\theta|\mathcal{X}_j)$ is high for $j\neq k$ and neither of the
additional MCMC method and IS correction works. On the other hand, if the scale factor is
too small, for example when $\lambda_k = 0.001, k= 1,\cdots, K$, we need more pairs
of sample points and corresponding logarithms of the probability density
function (pdf) of each subposterior to train a good approximation, even though the
subposteriors are more likely to overlap. In some cases where we can easily and cheaply obtain approximations
of the means and covariances of the true posterior and subposteriors (here,
$\lambda_k = 1, k=1, \cdots, K$), we can choose $\lambda_k$ such that
a chosen high probability posterior region is covered by one such region for the subposterior
$\pi_k$. Specifically, in a one-dimension case ($d=1$), let $\hat{\theta}$,
$\hat{\theta}_k$, $\hat{\sigma}$ and $\hat{\sigma}_k$ be the approximations of
the means and standard deviations of the true posterior and the subposterior
$\pi_k$, respectively. Denote $\delta_k = \max\{|\hat{\theta}_k - \hat{\theta}
-2\hat{\sigma}|,|\hat{\theta}_k - \hat{\theta} +2\hat{\sigma}|\}$ and choose
$\lambda_k = (\delta_k/\hat{\sigma}_k)^{-2}$. By Markov's inequality,
$\pi_k^{\lambda_k}$ covers most of a high probability region for the true posterior.
In high dimension cases, i.e., when $d > 1$, we can choose the scale factors as the
minimal ones according to each marginal component.\\
\\
\textbf{Remark 2:} \textbf{(Regression algorithms)} Considering the easy
implementation, strong learning ability of non-linearity and robustness of
random forests, we apply this modelling technique to learn cheapapproximations of the 
subposterior density function. Of course, other regression algorithms could be
chosen instead. However, when compared with random forests, these other machine learning
algorithms, such as support vector machines or neural networks, have a higher cost in terms of
hyper-parameters to set, while random forests are both easy to tune and relatively robust.
Besides, prediction by random forests is scalable, that is, given the training
set size $M$, the cost of predicting the output of a new input is just
$\mathcal{O}(\log(M))$. When we use these approximations of the subposteriors
$f_k$ to run an additional MCMC, the evaluation cost of a new proposal is of order
$\mathcal{O}(K\log(M))$, to compare with $\mathcal{O}(KM)$ in memory-based algorithms,
such as local linear/polynomial regression, nonparametric kernel density
estimation, Gaussian processes.
\\
\\
\textbf{Remark 3:} \textbf{(Combination of Importance Sampling)} For our second
method, we use importance sampling over each subposterior and combine them to
approximate the true posterior, without need to run an additional MCMC.
However, considering that the subposteriors are more spread out than the true
posterior, a large portion of their samples has extremely low weights and
need be discarded to achieve an better approximation of the true posterior. Besides,
in importance sampling, the proposal samples are independent. As a
result, in the MCMC stage of subposteriors, we thin out the accepted samples such
that they are approximately independent. More precisely, suppose
$\left\{\theta_1^k, \cdots, \theta_T^k\right\}$ are the well tuned samples of
$\pi_k^{\lambda_k}$, their weights are 
$$w_t^k \propto
\exp\left\{\sum_{j=1}^Kf_j(\theta_t^k) - \lambda_kf_k(\theta_t^k)\right\},
$$
for $t=1, \cdots, T$. Let $\sigma$ be a permutation of $\{1, \cdots, T\}$ such that
$w_{\sigma(1)}^k\geq w_{\sigma(2)}^k\geq\cdots\geq w_{\sigma(T)}^k$. According
to a pre-specified truncation probability $p$, we choose $i_k = \min\{i:
\sum_{\ell=1}^iw^k_{\sigma(\ell)} \geq p\}$ and treat $\tilde{\pi}_k =
\sum_{\ell=1}^{i_k}\tilde{w}_{\ell}^k\delta_{\theta_{\sigma{\ell}}^k}$, where
$\tilde{w}_{\ell}^k ={w_{\ell}^k}\big/{\sum_{r=1}^{i_k}w_r^k}$. Generally, we
can choose $p = 0.99, 0.999, 0.9999.$ Considering that the $i_k$'s are unequal
across $k=1, \cdots, K$ by stochasticity, some $\tilde{\pi}_k$'s have more atoms than average,
while others have fewer atoms. We weight the approximation $\tilde{\pi}_k$
proportional to their effective sample sizes (ESS)
\citep{liu2008monte}, that is, we approximate the true posterior with
\begin{equation*}
\hat{\pi} \propto \sum_{k=1}^KESS_k\tilde{\pi}_k,\quad ESS_k = \frac{i_k}{1+\mathbb{V}_k}
\end{equation*}
where $\mathbb{V}_k$ is the variance of $\{i_kw_{\sigma(1)}^k, \cdots, i_kw_{\sigma(i_k)}^k\}$.\\
\\
\textbf{Remark 4:} \textbf{(Computation complexity)} 
\begin{enumerate}
\item In the divide-and-conquer stage, the computation cost is
$\mathcal{O}(T_kN/K)$ on each computation node and we generate $T$ well-turned
samples points.
\item In the regression training stage, the cost for each random forest is $\mathcal{O}(T_k\log T_k)$
\item In the combination stage
   \begin{enumerate}
   \item To run an additional MCMC, the computation cost of a proposal point is $\mathcal{O}(K\log T_k)$.
   \item To run importance sampling, the cost of weighting all well-tuned
samples of all subposteriors on each computation node is $\mathcal{O}(KT\log T_k)$.
   \end{enumerate}
\end{enumerate}

\section{Numerical Experiments}
For simplicity of notations, we call the first method RF-MH and the second one
RF-IS. In this section, we apply our methods to several examples, from Gaussian
posterior to strongly non-Gaussian posteriors and to model misspecification
case.  We compare our methods with consensus Monte Carlo (CMC)
\cite{scott2016bayes}, nonparametric KDE (Nonpara)
\cite{neiswanger2013asymptotically} and Weierstrass sampler (Weierstrass)
\cite{wang2013parallelizing}.
\subsection{Bimodal Posterior}
In this example, the parameter $\theta = (\theta_1, \theta_2)$ drives the model
\begin{equation*}
\begin{split}
X\sim \frac{1}{2}\mathcal{N}(\theta_1, 2) + \frac{1}{2}\mathcal{N}(\theta_1+\theta_2, 2),
\end{split}
\end{equation*}
This toy example comes from \cite{welling2011bayesian}. We use the flat prior
$\pi_0(\theta_1, \theta_2)\propto 1$. We draw $n=200$ observations $\{X_1, X_2,
\cdots, X_{200}\}$ from the model with $\theta_1 = 0, \theta_2 = 1$. In order
to apply the parallel MCMC algorithms, we set $K = 10$ and $20$. There are two
configurations of the parameters corresponding to the observations, $(\theta_1,
\theta_2) = (0,1)$ and $(1,-1)$. When the observation size is moderate, both
configurations can be detected and the posterior is bimodal. While the size is
too large, Markov chain will be stuck at one mode, which will output a unimodal
posterior. In order to show that our proposed methods can handle non-Gaussian
posterior distributions, we set a moderate $n=200$. We generate 500,000 samples
for each subposterior, discard the first 100,000 points and thin out the rest per
100 sample points. In this example, we set $\lambda_k = 1, k=1, \cdots, K$ and
train each random forest with $10$ random partition trees on a random subset of
those sample sets with size 50,000. In Figure $1$, we  compare the contours of
the resulting pdfs derived from the samples for each method for $K=10$. It is clear that consensus
Monte Carlo could not capture the bimodal feature of the posterior, and
nonparametric KDE gives unequal weights for two modes, while our methods and
Weierstrass sampler perform very well.  In Figure $2$, we set $K = 20$, except
for our method, other methods deviate the true posteriors. In Table 1, we
compare the time budget of each method. In this example, the MCMC stage
dominates in the computation cost, all but the nonparametric KDE have the same
order of time budget. Since KDE is memory-based, its prediction cost is
$\mathcal{O}(T^2)$, while our method exhibits scalable prediction cost.
\begin{figure}[h]
\centering
\includegraphics[width=14cm]{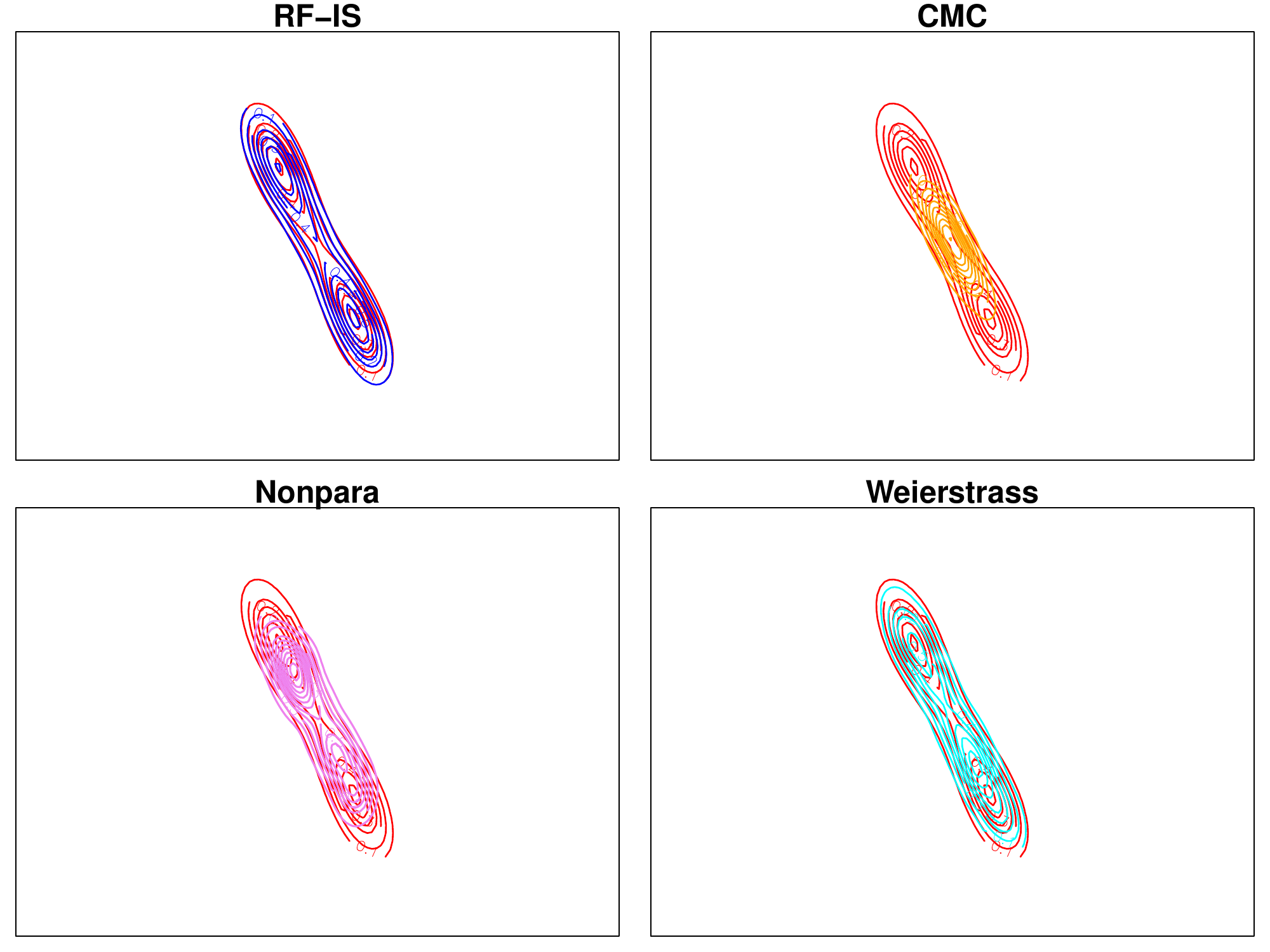}
\caption{Example 1: Comparisons of the contours of true posterior (red), RF-IS(blue), consensus Monte Carlo (orange), KDE (violet) and Weierstrass sampler (cyan) for $K=10$}
\end{figure}
\begin{figure}[h]
\centering
\includegraphics[width=14cm]{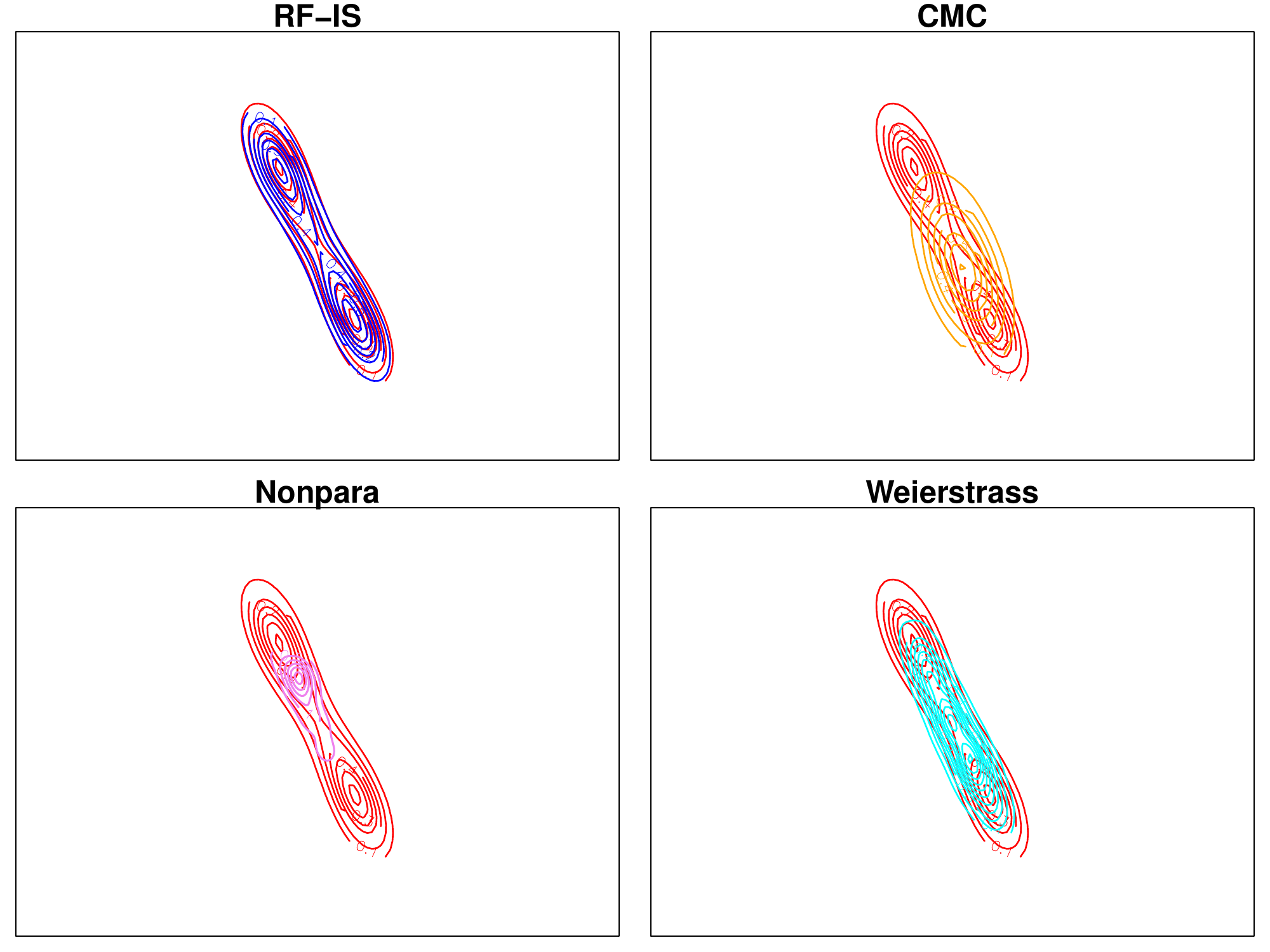}
\caption{Example 1: Comparisons of the contours of true posterior (red), RF-IS(blue), consensus Monte Carlo (orange), KDE (violet) and Weierstrass sampler (cyan) for $K=20$}
\end{figure}
\begin {table}[H]
\begin{center}
\begin{tabular}{| c | c | c | c | c|}
  \hline 
  \multirow{1}{*} {K} &RF-IS & CMC & Nonpara & Weierstrass \\ 
  \hline
  \multirow{4}*{$K=10$}                                                                   
               &MCMC   \hspace{.1cm}    $54.8$& MCMC   \hspace{.1cm}    $54.8$ & MCMC   \hspace{0.1cm}   $54.8$&MCMC   \hspace{0.1cm}   $54.8$\\
               &Training  \hspace{.1cm} $2.8$&  & &\\
               &Weighting\hspace{.1cm}  $0.3$& Combination \hspace{.1cm}    $0.1$&Combination \hspace{.1cm}    $109.3$&Combination \hspace{.1cm}    $0.8$\\
               \cline{2-1}
               \cline{3-1}
               \cline{4-1}
               \cline{5-1}
               &Total \hspace{.1cm} $57.9$&Total \hspace{.1cm} $54.9$&Total \hspace{.1cm} $164.1$&Total \hspace{.1cm} $55.6$\\
  \hline
  \multirow{4}*{$K=20$}                                                                   
               &MCMC   \hspace{.1cm}    $52.8$& MCMC   \hspace{.1cm}     $52.8$ & MCMC   \hspace{0.1cm}     $52.8$ &MCMC   \hspace{0.1cm}     $52.8$\\
               &Training      \hspace{.1cm}    $2.8$&  & &\\
               &Weighting   \hspace{.1cm}   $0.6$& Combination \hspace{.1cm}    $0.2$&Combination \hspace{.1cm}    $403.2$&Combination \hspace{.1cm}    $1.6$\\
               \cline{2-1}
               \cline{3-1}
               \cline{4-1}
               \cline{5-1}
               &Total \hspace{.1cm} $56.2$&Total \hspace{.1cm} $53.0$&Total \hspace{.1cm} $456.0$&Total \hspace{.1cm} $54.4$\\
  \hline
\end{tabular}
\caption {Time budget of Example 1 (in seconds).}
\end{center}
\end {table}
\subsection{Moon-shaped Posterior}
This is a toy example which is unidentifiable in the parameters.  Denote
$\pmb{\theta} = (\theta_1, \theta_2) \in [0,\infty)^2$ as parameters, the
observations, $X_1, \cdots, X_N$, given the parameter $\pmb{\theta}$, are
generated from
\begin{equation*}
X \sim \mathcal{N}(\sqrt{\theta_1}+\sqrt{\theta_2},2), \quad \theta_1 \geq 0, \theta_2 \geq 0
\end{equation*} 
In this model, in light of the non-identifiability, the posterior is
moon-shaped and strongly non-Gaussian. In our experiment, we generate
observations from standard normal distribution, $\mathcal{N}(0,1)$, set $N=
1000, K=10,  20, \lambda_k = 1, k=1, \cdots, K$ and draw 500,000 sample points
from each subposterior, burn the first 100,000 ones and thin out the rest per 100
points. Using large proposal points of each subposterior, we train each random
forest with $10$ random partition trees on random subset of the samples set
with size 50,000. Figure $3$ compares the performances of each method for
$K=10$ case. In this strongly non-Gaussian case, consensus Monte Carlo and
nonparametric KDE produce Gaussian posteriors. To some extent the Weierstrass
sampler is able to learn this non-Gaussian posterior, but our method has the
best performance thanks to its random forest construct. In Figure 4, $K=20$ and
we can check that the Weierstrass sampler becomes worse while our method still
approximates closely the target. In Table 2, we present the time budget of each
method. We can see there that nonparametric KDE is the most costly solution,
since it is memory-based and its cost of a new prediction is
$\mathcal{O}(T^2)$. Since the MCMC stage over subsets dominates the computing
budget, our methods has the same order of time budget as consensus Monte Carlo
and Weierstrass samplers. 
\begin{figure}[h]
\centering
\includegraphics[width=14cm]{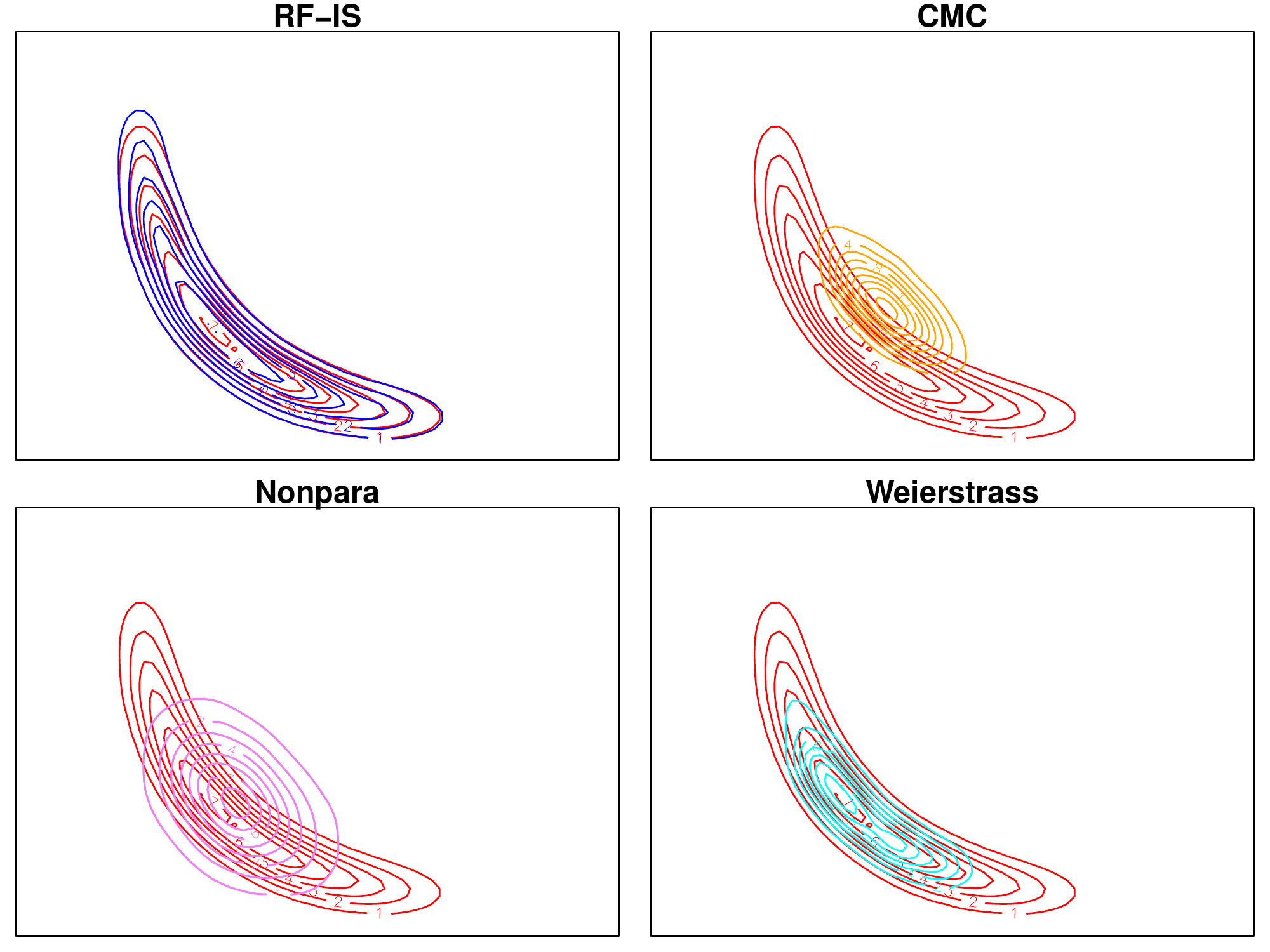}
\caption{Example 2: Comparisons of the contours of true posterior (red), RF-IS(blue), consensus Monte Carlo (orange), KDE (violet) and Weierstrass sampler (cyan) for $K=10$}
\end{figure}
\begin{figure}[h]
\centering
\includegraphics[width=14cm]{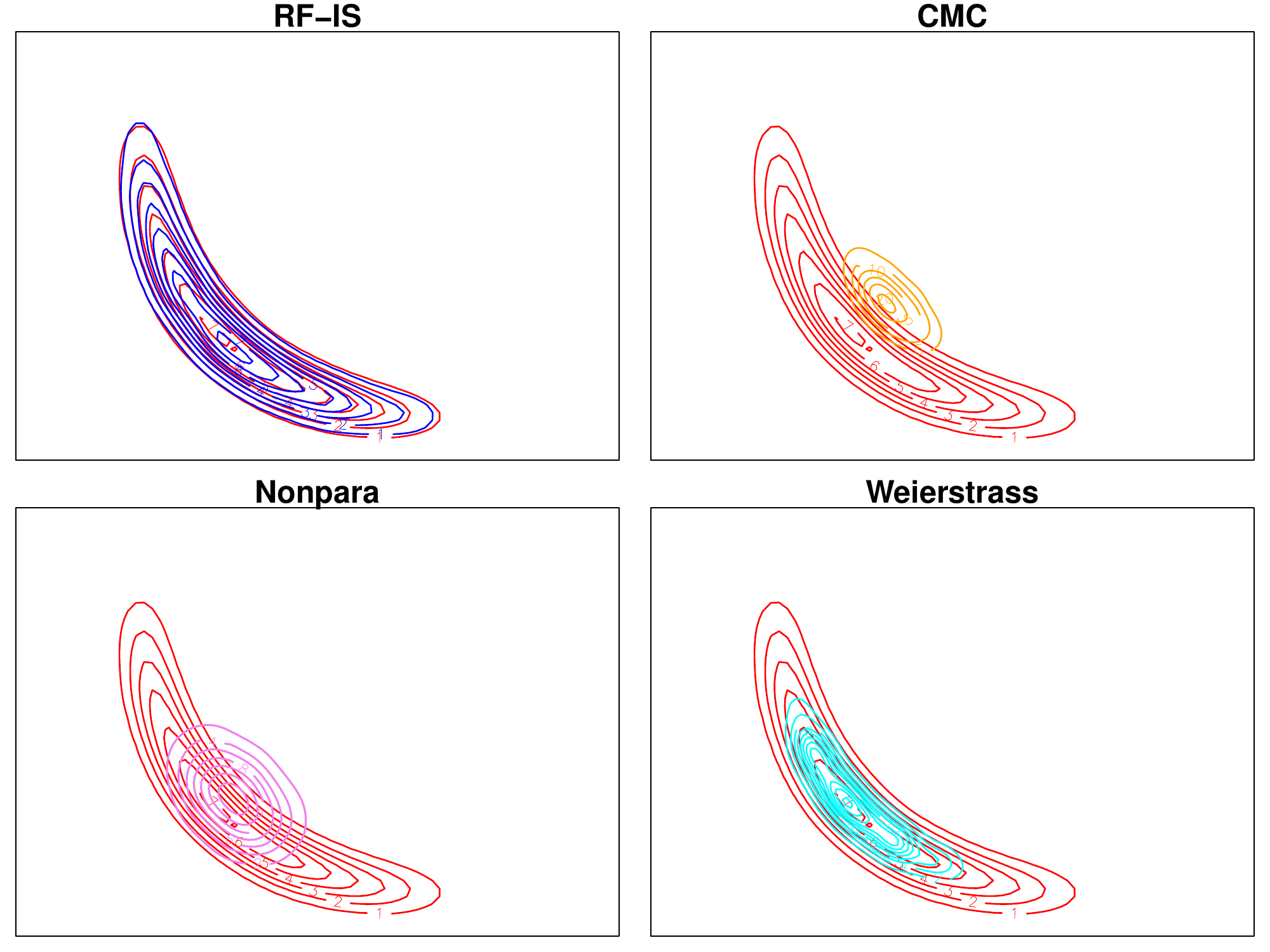}
\caption{Example 2: Comparisons of the contours of true posterior (red), RF-IS(blue), consensus Monte Carlo (orange), KDE (violet) and Weierstrass sampler (cyan) for $K=20$}
\end{figure}
\begin {table}[H]
\begin{center}
\begin{tabular}{| c | c | c | c | c|}
  \hline 
  \multirow{1}{*} {K} &RF-IS & CMC & Nonpara & Weierstrass \\ 
  \hline
  \multirow{4}*{$K=10$}                                                                   
               &MCMC   \hspace{.1cm}    $67.6$& MCMC   \hspace{.1cm}    $67.6$ & MCMC   \hspace{0.1cm}   $67.6$&MCMC   \hspace{0.1cm}   $67.6$\\
               &Training  \hspace{.1cm} $2.8$&  & &\\
               &Weighting\hspace{.1cm}  $0.3$& Combination \hspace{.1cm}    $0.1$&Combination \hspace{.1cm}    $101.7$&Combination \hspace{.1cm}    $0.7$\\
               \cline{2-1}
               \cline{3-1}
               \cline{4-1}
               \cline{5-1}
               &Total \hspace{.1cm} $70.7$&Total \hspace{.1cm} $67.7$&Total \hspace{.1cm} $169.3$&Total \hspace{.1cm} $68.3$\\
  \hline
  \multirow{4}*{$K=20$}                                                                   
               &MCMC   \hspace{.1cm}    $58.7$& MCMC   \hspace{.1cm}     $58.7$ & MCMC   \hspace{0.1cm}     $58.7$ &MCMC   \hspace{0.1cm}     $58.7$\\
               &Training      \hspace{.1cm}    $2.7$&  & &\\
               &Weighting   \hspace{.1cm}   $0.7$& Combination \hspace{.1cm}    $0.2$&Combination \hspace{.1cm}    $371.5$&Combination \hspace{.1cm}    $1.5$\\
               \cline{2-1}
               \cline{3-1}
               \cline{4-1}
               \cline{5-1}
               &Total \hspace{.1cm} $62.1$&Total \hspace{.1cm} $58.9$&Total \hspace{.1cm} $430.2$&Total \hspace{.1cm} $60.2$\\
  \hline
\end{tabular}
\caption {Time budget of Example 2 (in seconds).}
\end{center}
\end {table}
\subsection{Misspecification Example}
This example is borrowed from \cite{bardenet2015markov} and is to show that
with a suitable selection of the scale factors, $\lambda_k$, our methods are
robust to misspecification of models. In the example, the model is
\begin{equation*}
X \sim \mathcal{N}(\mu,\sigma^2),
\end{equation*}
the parameter is $\pmb{\theta} = (\mu, \sigma^2)$. In the experiments, we
generate two data sets, each having $N=10000$ points $X_1, \cdots, X_{10000}$,
from log-normal distribution, $\mathcal{LN}(0,1)$  and standard normal
distribution, $\mathcal{N}(0,1)$, respectively, and choose the flat prior,
$p(\mu,\sigma^2)\propto 1$. Let $K=10$ and choosing $\lambda_k$ from the method
in Remark 1.  In more details, by the maximum likelihood estimator, we have, over
the whole data set, 
\begin{equation*}
\hat{\pmb{\theta}} = \left({\sum_{i=1}^Nx_i}\big/{N}, {\sum_{i=1}^N(x_i-\bar{x})^2}\big/{N}\right)
\qquad\qquad
\hat{\pmb{\sigma}}_{\pmb{\theta}} = \left(\sqrt{{\hat{\theta}_2}\big/{N}}, \sqrt{{2\hat{\theta}_2^2}\big/{N}}\right)
\end{equation*}
For each subset $\mathcal{X}_k$, we can obtain $\hat{\pmb{\theta}}^{(k)}$ and $\hat{\pmb{\sigma}}_{\pmb{\theta}^{(k)}}$ similarly. Set 
\begin{equation*}
\lambda_{k,1} = \left(\frac{\max\{|\hat{\theta}^{(k)}_1 - \hat{\theta}_1-2\hat{\sigma}_{\theta_1}|, |\hat{\theta}^{(k)}_1 - \hat{\theta}_1+2\hat{\sigma}_{\theta_1}|\}}{\hat{\sigma}_{\theta^{(k)}_1}}\right)^{-2}
\end{equation*}
\begin{equation*}
\lambda_{k,2} = \left(\frac{\max\{|\hat{\theta}^{(k)}_2 - \hat{\theta}_2-2\hat{\sigma}_{\theta_2}|, |\hat{\theta}^{(k)}_2 - \hat{\theta}_2+2\hat{\sigma}_{\theta_2}|\}}{\hat{\sigma}_{\theta^{(k)}_2}}\right)^{-2}
\end{equation*}
and $\lambda_k = \min\{\lambda_{k,1},\lambda_{k,2}\}$.
We generate 500,000 sample points from each subposterior, discard the first 100,00
values and thin out the rest per 100 points for Gaussian case and per 10 points for
Log-Normal case. Each random forest is trained with 10 random partition trees.
Figure $5$ and Figure $6$ present the performance of each method under the
normal distribution and log-normal distribution respectively. In each case,
samples of nonparametric KDE are far away from the true posterior, as some
subposteriors do not cover the high probability region of true posterior. In
the normal case, RF-IS, consensus Monte Carlo and Weierstrass sampler all perform
very well. However, only our method appears to be robust to model misspecification, the
other ones deviating from the true posteriors to some extent in the log-normal case. In
Table 3, we compare the time budget for each method and we can spot that
the nonparametric KDE is the most costly in the log-normal case as the size of samples
from each subposterior is quite large.
\begin{figure}[!h]
\centering
\includegraphics[width=14cm]{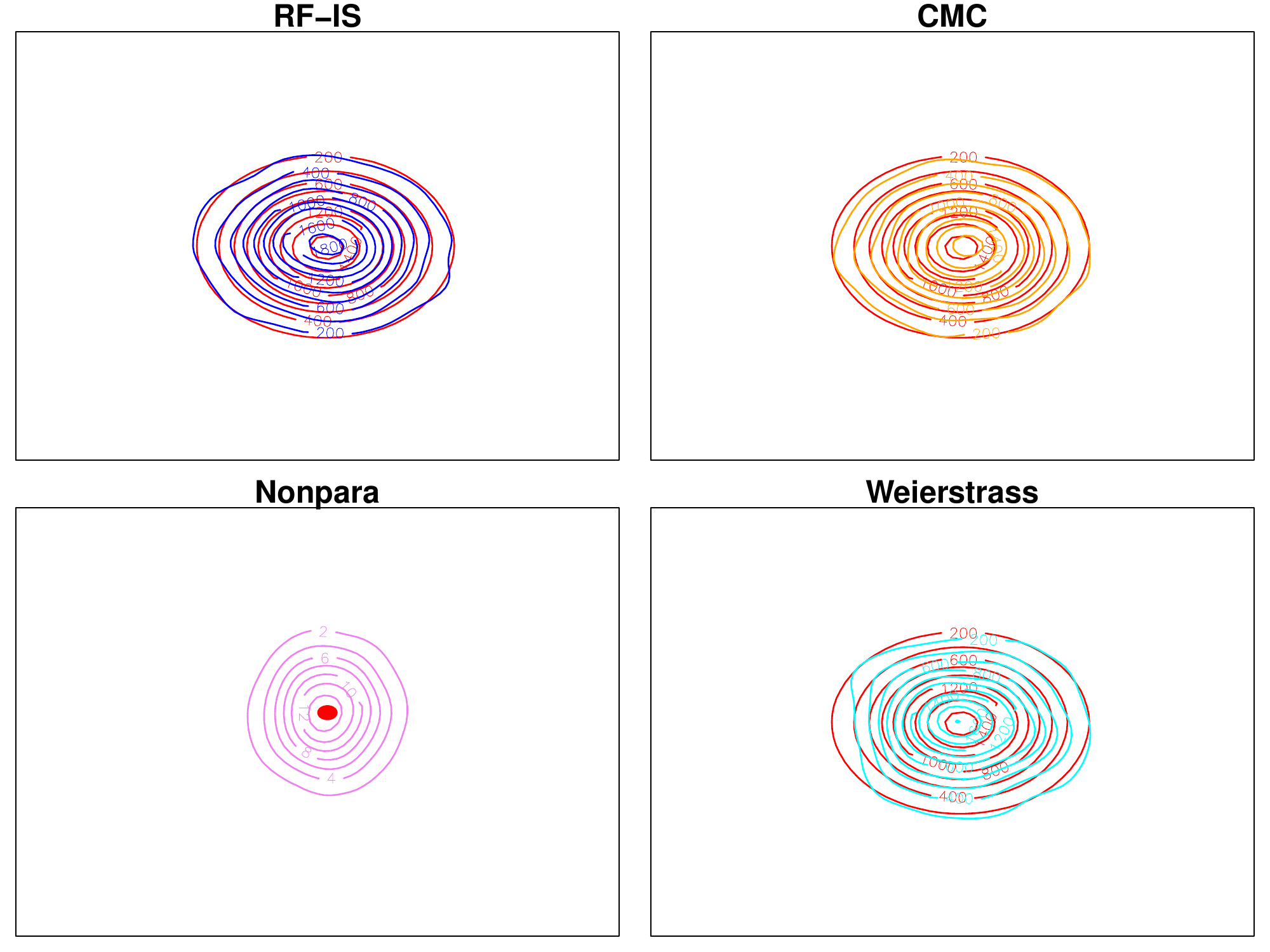}
\caption{Example 3: For normal observations, comparisons of the contours of true posterior (red), RF-IS(blue), consensus Monte Carlo (orange), KDE (violet) and Weierstrass sampler (cyan). }
\end{figure}
\begin{figure}[!h]
\centering
\includegraphics[width=14cm]{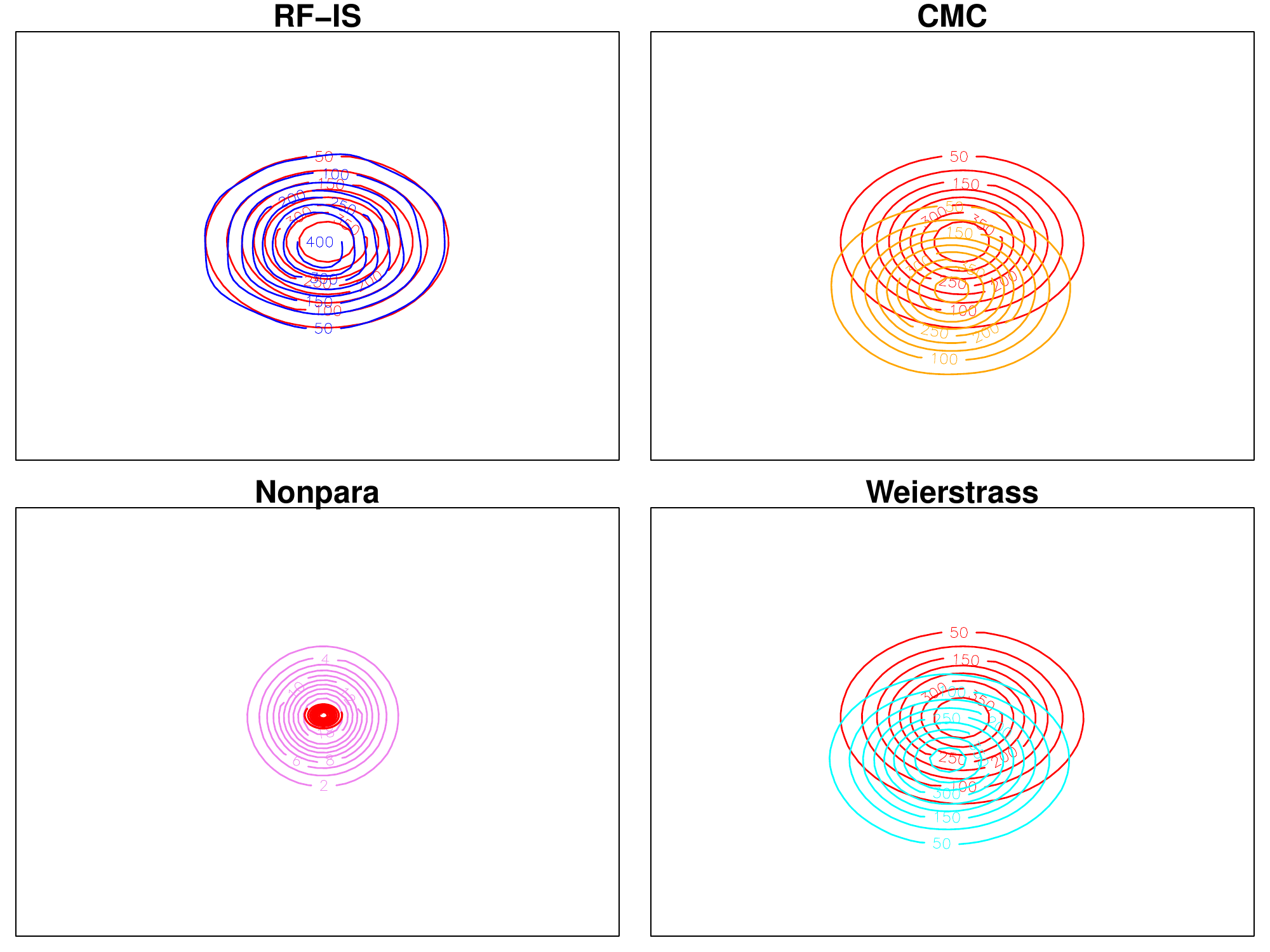}
\caption{Example 3: For log-normal observations, comparisons of the contours of true posterior (red), RF-IS(blue), consensus Monte Carlo (orange), KDE (violet) and Weierstrass sampler (cyan). }
\end{figure}
\begin {table}[H]
\begin{center}
\begin{tabular}{| c | c | c | c | c|}
  \hline 
  \multirow{1}{*} {Model} &RF-IS & CMC & Nonpara & Weierstrass \\ 
  \hline
  \multirow{4}*{$\mathcal{N}(0,1)$}                                                                   
               &MCMC   \hspace{.1cm}    $63.9$& MCMC   \hspace{.1cm}    $65.2$ & MCMC   \hspace{0.1cm}   $65.2$&MCMC   \hspace{0.1cm}   $65.2$\\
               &Training  \hspace{.1cm} $117.4$&  & &\\
               &Weighting\hspace{.1cm}  $1.1$& Combination \hspace{.1cm}    $0.1$&Combination \hspace{.1cm}    $99.6$&Combination \hspace{.1cm}    $0.8$\\
               \cline{2-1}
               \cline{3-1}
               \cline{4-1}
               \cline{5-1}
               &Total \hspace{.1cm} $182.4$&Total \hspace{.1cm} $65.3$&Total \hspace{.1cm} $164.8$&Total \hspace{.1cm} $66.0$\\
  \hline
  \multirow{4}*{$\mathcal{LN}(0,1)$}                                                                   
               &MCMC   \hspace{.1cm}    $67.4$& MCMC   \hspace{.1cm}     $69.2$ & MCMC   \hspace{0.1cm}     $69.2$ &MCMC   \hspace{0.1cm}     $69.2$\\
               &Training      \hspace{.1cm}    $119.4$&  & &\\
               &Weighting   \hspace{.1cm}   $5.3$& Combination \hspace{.1cm}    $0.9$&Combination \hspace{.1cm}    $1114.4$&Combination \hspace{.1cm}    $7.5$\\
               \cline{2-1}
               \cline{3-1}
               \cline{4-1}
               \cline{5-1}
               &Total \hspace{.1cm} $192.1$&Total \hspace{.1cm} $70.1$&Total \hspace{.1cm} $1183.6$&Total \hspace{.1cm} $76.7$\\
  \hline
\end{tabular}
\caption {Time budget of Example 3 (in seconds).}
\end{center}
\end {table}

\section{Conclusion}
Our proposal is therefore to combine divide-and-conquer MCMC methods, random forests and
importance sampling to scale MCMC algorithms. Unlike the existing
divide-and-conquer MCMC methods, we propose to scale the partial posterior with
factors which are not necessarily $1$ or the selected number of subsets or to
be equal with one another. Given suitable scale factors, we can achieve overlapping
subposteriors, a feature that is of the highest importance in the combination stage.
Considering its strong non-linear learning ability, an easy implementation and
a scalable prediction cost, a method based on random forests embedded in a divide-and-conquer
MCMC framework delivers cheap and robust approximations of the subposteriors.
Overall, our numerical experiments achieve good performance, from
Gaussian cases to strongly non-Gaussian cases, which are further to model
misspecification, exhibiting to some extent limitations of existing divid-and-conquer
scalable MCMC methods.\\

The main limitation of our method is the curse of dimensionality in random
forest training. In high dimensional parameter spaces, it requires more sample
points to train each random forest learner. The second shortcoming is the 
necessary selection of scale factors. We can offer no generic method to tune them, except in
some cases where we can obtain cheap estimations of the means and covariances
of the true posteriors and the subposteriors.\\

In fact, if we can roughly detect a high probability region, $E$, for the
posterior, we can discard the stage of running MCMC over subsets, instead
generating points from $E$ uniformly and training the random forests with these
points and their corresponding pdf values for each subposterior, respectively.
Besides, the result of our methods can also be used to produce such $E$. When considering
the practical implementation of this extension, since RF-MH need store $K$ random forests at each
iteration in order to predict the likelihood value, it proves more costly than
RF-IS, even though the computation complexity is the same as RF-IS in theory.

\end{document}